\documentclass[11pt, fceqn]{article}
 \usepackage{amsfonts}
 \usepackage{flafter}
 \usepackage{amssymb,graphics,graphicx,subfigure,caption2,rotating}
 \usepackage{amsmath, latexsym}
\usepackage[amsmath,thmmarks]{ntheorem}
 \usepackage[top=3.5cm]{geometry}
\begin{document}

\date{\today}
\title{Quantum relaxed row and column iteration methods based on block-encoding}
\author{Xiao-Qi Liu$^1$, Jing Wang$^1$, Ming Li$^1$$^2$\thanks{Electronic address: liming@upc.edu.cn.}, Shu-Qian Shen$^1$, Weiguo Li$^1$, Shao-Ming Fei$^2$$^3$\\{\footnotesize{\it $^1$College of Science, China University of Petroleum, 266580 Qingdao, P.R. China.}}\\
{\footnotesize{\it $^2$Max-Planck-Institute for Mathematics in Sciences, 04103 Leipzig, Germany.}}\\
{\footnotesize{\it $^3$School of Mathematical Science, Capital Normal University, 100048, Beijing, China.}}}\maketitle

\begin{abstract}
Iteration method is commonly used in solving linear systems of equations. We present quantum algorithms for the relaxed row and column iteration methods by constructing unitary matrices in the iterative processes, which generalize row and column iteration methods to solve linear systems
on a quantum computer. Comparing with the conventional row and column iteration methods, the convergence accelerates when appropriate parameters are chosen. Once the quantum states are efficiently prepared, the complexity of our relaxed row and column methods is improved exponentially and is linear with the number of the iteration steps. In addition, phase estimations and Hamiltonian simulations are not required in these algorithms.
\end{abstract}

\maketitle

\section{Introduction}

Based on the state superposition and quantum entanglement, quantum algorithms can solve problems faster than classical algorithms, and even handle problems beyond the power of classical computers \cite{Shor,Grover}. A number of quantum algorithms have been presented which are superior to the classical counterparts. The HHL algorithm proposed in \cite{HHL}
solves set of linear equations, and accelerates exponentially over classical algorithms under certain conditions. In the era of big data, quantum machine learning algorithms \cite{QML} have provided new solutions to the efficiency problems caused by the large amount and high-dimensional data. Quantum versions of machine learning algorithms such as quantum support vector machines \cite{QSVM} and quantum neural networks \cite{FNN} have been derived. Furthermore, it is generally believed that the quantum computing community has entered the computing era of noisy intermedial-scale quantum (NISQ) \cite{NISQ} and variational quantum algorithms \cite{VQA} have been widely used for NISQ devices.

Since many problems in engineering technology and natural science can eventually be transformed into solving linear equation sets, it is of great significance to solve linear equations effectively. There are two common methods to solve linear equations: direct method and iterative method. The direct method, such as Gaussian elimination method, seeks the exact solution of linear equations by finite step operations of ``elimination" and ``back substitution". The iteration method constructs iterative formula, the finite step successive iterative approximation, to get the approximate solution of the equations. Generally speaking, direct method offers an outstanding performance in low order linear equations. While dealing with large scale sparse matrices, the iteration method provides high efficiency.

Both row and column iteration methods are widely used iteration methods. In the early 20th century, Kaczmarz proposed the illustrious row method \cite{Kaczmarz} to solve linear equations. Bender and Herman rediscovered Kaczmarz method in the field of image reconstruction, which is called algebraic reconstruction technology \cite{Gordon}. The Kaczmarz method is the most commonly used method in solving large coefficient matrices because of its efficiency and simplicity. The column iterative method is the coordinate descent method. Inspired by the randomized Kaczmarz method, Leventhal and Lewis \cite{CD} proposed the randomized coordinate descent method and verified that the linear convergence rate is ideal. The column iteration method seeks a least squares solution generally and the Kaczmarz method calculates a minimum norm solution.

For the problem with size $n$ and iteration steps $T$, the complexity of the classical iteration method is polynomial in $n$ and linear in $T$. For most of the standard quantum iterative algorithms \cite{iteration1,iteration2}, they are not superior to the classical iterative algorithms neither in $n$ nor $T$. In addition, the previous quantum algorithms of linear equation solution have different assumptions \cite{HHL,CKS,WZP,CGJ}. Based on the idea of block-encoding \cite{CGJ} Shao and Xiang proposed a new method for solving liner systems and realized exponential acceleration \cite{Shao,Zuo}.

In this paper, based on the idea of block-encoding we present the relaxed row and column iteration quantum algorithms. Our algorithms can accelerate the convergence of iteration by selecting appropriate relaxation factors.

\section{Relaxed row and column iteration quantum algorithms}

Consider a linear system of equations given by $Ax=b$, where $A$ is an $n\times n$ matrix. Denote $a_{t}={(a_{t1},a_{t2},...,a_{tn})}^{T}$. We have $a_{t}^T x=b_{t}$, $t=1,2,\cdots,n$. The Kaczmarz method starts from the initial point, and iterates several times to get an approximate solution by projecting onto the selected hyperplane each time.
The iterative formula of Kaczmarz is given by
\begin{equation*}
x_{k+1}=x_{k}+\frac{b_{t_{k}}-{a_{t_{k}}}^T x_{k}}{{a_{t_{k}}}^T a_{t_{k}}}a_{t_{k}},\quad k=0,1,2,\cdots,\quad t_k=1,2,\cdots,n.
\end{equation*}
The relaxed Kaczmarz method \cite{relaxed kaczmarz} is to add a relaxation factor $\lambda$ in each step, with the iterative formula given by
\begin{equation}
\label{lambda}
x_{k+1}=x_{k}+\lambda_{k}\frac{b_{t_{k}}-{a_{t_{k}}}^T x_{k}}{{a_{t_{k}}}^T a_{t_{k}}}a_{t_{k}},\quad k=0,1,2,\cdots,\quad t_k=1,2,\cdots,n.
\end{equation}

Let $c_{t}$ be the $t$th column of $A$, i.e. $A=(c_{1},c_{2},\cdots,c_{n})$ and $e_{t_k}=(0,\cdots,0,e_{t_k mod n}=1,0,\cdots,0)^{T}$. The column iterative formulae are given by
\begin{eqnarray*}
x_{k+1}&=&x_{k}+\frac {c_{t_{k}}^T r_{k}}{c_{t_{k}}^T c_{t_{k}}}e_{t_{k}},\\
r_{k+1}&=&r_{k}-\frac {c_{t_{k}}c_{t_{k}}^T}{c_{t_{k}}^T c_{t_{k}}}r_{k},\quad k=0,1,2,\cdots,\quad t_k=1,2,\cdots,n.
\end{eqnarray*}
And the classical iterative formulas of the relaxed column iteration \cite{relaxed cd1,relaxed cd2} are given by
\begin{eqnarray}
\label{omega1}
&&x_{k+1}=x_{k}+\omega_{k}\frac {c_{t_{k}}^T r_{k}}{c_{t_{k}}^T c_{t_{k}}}e_{t_{k}},\\
\label{omega2}
&&r_{k+1}=r_{k}-\omega_{k}\frac {c_{t_{k}}c_{t_{k}}^T}{c_{t_{k}}^T c_{t_{k}}}r_{k},\quad k=0,1,2,\cdots,\quad t_k=1,2,\cdots,n.
\end{eqnarray}
It can be seen from iterative formulae that when $\lambda_{k}\equiv1$ $(\omega_{k}\equiv1)$ for all $k$, the above formulae reduce to the ones in the classical row and column iteration methods. Whitney and Meany \cite{Whitney} proved that the relaxed Kaczmarz method converges provided that the relaxation parameters are within $[0, 2]$ and $\lambda_k\rightarrow 0$, when $k\rightarrow \infty$.

Inspired by the relaxation strategy, if the step size of a point in the projection direction becomes longer or shorter, it will be more beneficial to the projection of the next hyperplane, see FIG. 1. Therefore, choosing an expedient relaxation factor will greatly accelerate the convergence of iteration.
\begin{figure}
\centering
\includegraphics [scale=0.25]{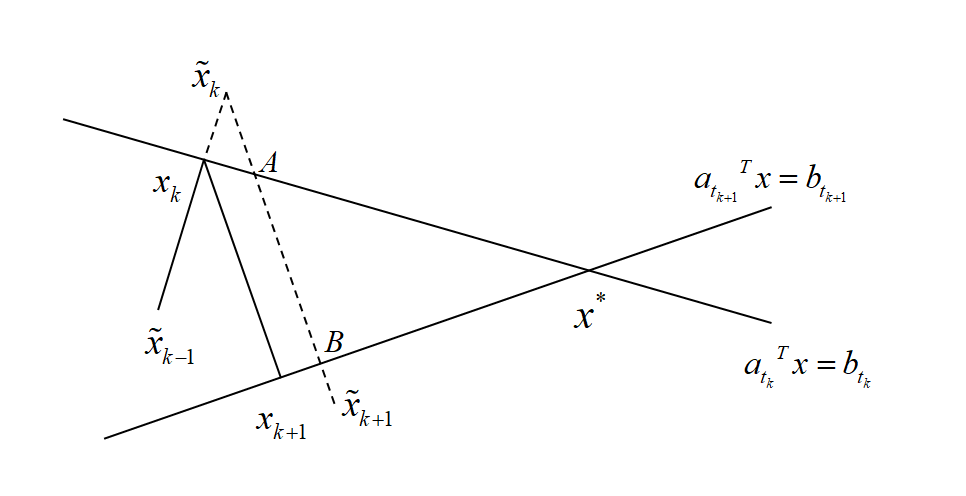}
\caption{$\widetilde{x}_{k+1}$ is the projection of $\widetilde{x}_{k}$ in the relaxed Kaczmarz algorithm at the $k+1$ iteration. Compared with the $x_{k+1}$ of the Kaczmarz algorithm at the $k+1$ iteration, $\widetilde{x}_{k+1}$ is closer to the solution $x^{\ast}$ of the equations.}
\end{figure}

\subsection{Quantum relaxed Kaczmarz algorithm}

We first give the quantum relaxed Kaczmarz algorithm. Without loss of generality, we can assume that for all $t$, $||a_{t}||=1$ in the classical iteration formula of Kaczmarz algorithm given in $(\ref{lambda})$. Then the formula is simplified to be
\begin{equation}
\label{informaion}
x_{k+1}=x_{k}+\lambda_{k}(b_{t_{k}}-{a_{t_{k}}}^T x_{k})a_{t_{k}}.
\end{equation}
For preparing quantum state $|a_t\rangle$, we assume the existence of unitary matrix $V_t$ such that $V_t|0\rangle=|a_t\rangle$, $t=1,2,...,n$, for instance, by QRAM\cite{qram}. For some special cases\cite{cases}, the assumption of QRAM is unnecessary. The quantum state corresponding to the iterative formula
is given by
\begin{eqnarray*}
|x_{k+1}\rangle\varpropto||x_{k}||(I_{n}-\lambda_{k}|a_{t_{k}}\rangle \langle a_{t_{k}}|)|x_{k+1} \rangle+\lambda_{k}b_{t_{k}}|a_{t_{k}}\rangle.
\end{eqnarray*}

We prepare the state $|\varphi\rangle=\beta|0\rangle|X_{k}\rangle+\gamma|1\rangle|0\rangle|0\rangle|a_{t_{k}}\rangle$, where
\begin{equation*}
|X_{k}\rangle=\sqrt{p}|0\rangle|0\rangle|x_{k}\rangle+\sqrt{1-p}|1\rangle|0\rangle|\cdots\rangle,
\end{equation*}
with $\beta^2+\gamma^2=1$.
Let $SWAP_{i,j}$ be the unitary operator that swaps the $i$th and $j$th qubits. Applying $SWAP_{1,2}$ to $|\varphi\rangle$, we obtain
\begin{equation*}
|0\rangle\otimes(\beta\sqrt{p}|0\rangle|0\rangle |x_{k}\rangle+\gamma|1\rangle|0\rangle|a_{t_{k}}\rangle)+\beta\sqrt{1-p}|1\rangle|0\rangle|0\rangle|\cdots\rangle.
\end{equation*}
Set
\[
U_{t_{k}}=\begin{bmatrix}
I_{n}-\lambda_{k}|a_{t_{k}}\rangle \langle a_{t_{k}}| & \sqrt{2\lambda_{k}(1-\lambda_{k})}|a_{t_{k}}\rangle \langle a_{t_{k}}| & \lambda_{k}| a_{t_{k}} \rangle \langle a_{t_{k}} | & 0\\
\sqrt{2\lambda_{k}(1-\lambda_{k})}|a_{t_{k}}\rangle \langle a_{t_{k}}| & 2\lambda_{k}| a_{t_{k}}\rangle \langle a_{t_{k}} |-I_{n} & -\sqrt{2\lambda_{k}(1-\lambda_{k})}| a_{t_{k}}\rangle \langle  a_{t_{k}} | & 0\\
\lambda_{k}|a_{t_{k}} \rangle \langle a_{t_{k}}| & -\sqrt{2\lambda_{k}(1-\lambda_{k})}|a_{t_{k}}\rangle \langle a_{t_{k}} | & I_{n}-\lambda_{k}| a_{t_{k}}\rangle \langle a_{t_{k}} | & 0\\
0&0&0&I_{n}
\end{bmatrix}.
\]
Applying $U_{t_{k}}$ to $\beta\sqrt{p}|0 \rangle|0 \rangle |x_{k} \rangle+\gamma|1 \rangle| 0 \rangle| a_{t_{k}} \rangle$, one gets
\begin{align*}
U_{t_{k}}&(\beta\sqrt{p}| 0 \rangle| 0 \rangle |x_{k}\rangle+\gamma| 1 \rangle| 0 \rangle| a_{t_{k}} \rangle)\\
&=| 00 \rangle\otimes(\beta\sqrt{p}(I_{n}-\lambda_{k}| a_{t_{k}} \rangle \langle  a_{t_{k}} |)| x_{k} \rangle+\gamma\lambda_{k}| a_{t_{k}}\rangle)+| 0^\bot \rangle^{\otimes 2}| \cdots \rangle.
\end{align*}

The information contained in the relaxed Kaczmarz iterative formula $(\ref{informaion})$ is proportionately included in the first term of the above quantum state, by selecting the appropriate parameters $\beta=||x_k||\delta$ and $\gamma=b_t\sqrt p\delta$ for some $\delta$ to ensure that $\beta^2+\gamma^2=1$ \cite{Shao}.
Note that $| X_{k} \rangle$ and $| a_{t_{k}} \rangle$ can be produced by unitary operations. From the state $(\beta_{t_k}| 0 \rangle+\gamma_{t_k}| 1 \rangle)| 0 \rangle$, we obtain $| Y_{k} \rangle=\beta_{t_k}| 0 \rangle| X_{k} \rangle+\gamma_{t_k}| 1 \rangle| 0 \rangle^{\otimes 3k}| 00 \rangle| a_{t_{k}} \rangle$.
Then we apply $SWAP_{1,3k+2}$ to $| Y_{k} \rangle$, and get $| 0 \rangle^{\otimes (3k+1)}\otimes (\beta_{t_k}\frac{||x_{k}||}{v_{k}}| 00 \rangle| x_{k} \rangle+\gamma_{t_k}| 10 \rangle| a_{t_{k}} \rangle)$. In step 3, we obtain
\begin{align*}
|\widetilde{x}_{k+1}\rangle=&I_2^{\otimes(3k+1)}\otimes U_{t_{k}}(| 0 \rangle^{\otimes (3k+1)}\otimes (\beta_{t_k}\frac{||x_{k}||}{v_{k}}| 00 \rangle| x_{k} \rangle+\gamma_{t_k}| 10 \rangle| a_{t_{k}} \rangle)+| 0^\bot \rangle^{\otimes 3(k+1)}| \cdots \rangle)\\
=&\frac{\beta_{t_k}}{v_{k}}| 0 \rangle^{\otimes (3k+1)}\otimes| 00 \rangle\otimes(||x_{k}||(I_{n}-\lambda_{k}| a_{t_{k}} \rangle \langle  a_{t_{k}} |)| x_{k} \rangle+\lambda_{k}b_{t_k}| a_{t_{k}}\rangle)+| 0^\bot \rangle^{\otimes 3(k+1)}| \cdots \rangle\\
=&\frac{\beta_{t_k}}{v_{k}}| 0 \rangle^{\otimes 3(k+1)}\otimes(||x_{k}||(I_{n}-\lambda_{k}| a_{t_{k}} \rangle \langle  a_{t_{k}} |)| x_{k} \rangle+\lambda_{k}b_{t_k}| a_{t_{k}}\rangle)+| 0^\bot \rangle^{\otimes 3(k+1)}| \cdots \rangle\\
=&\frac{||x_{k+1}||}{v_{k+1}}| 0 \rangle^{\otimes 3(k+1)}| x_{k+1} \rangle+| 0^\bot \rangle^{\otimes 3(k+1)}| \cdots \rangle,
\end{align*}
where we have used the fact that $\gamma_{t_k}=\frac{\beta_{t_k}b_{t_k}}{v_{k}}$ and $\frac{\beta_{t_k}}{v_{k}}=\frac{1}{v_{k+1}}$.
Finally, by applying $SWAP_{1,3(k+1)+1}SWAP_{2,3(k+1)+2}$ to $| 00 \rangle|\widetilde{x}_{k+1}\rangle$, we obtain
\begin{equation*}
| X_{k+1} \rangle=\frac{||x_{k+1}||}{v_{k+1}}| 0 \rangle^{\otimes 3(k+1)}| 00 \rangle| x_{k+1} \rangle+| 0^\bot \rangle^{\otimes 3(k+1)+2}| \cdots \rangle.
\end{equation*}

The detailed flow of the quantum relaxed Kaczmarz algorithm is given in Table I.

From the algorithm, we can see that all the information of $| x_{k} \rangle$ can be obtained from the quantum state $| X_{k} \rangle$. In addition, when $\lambda_{k}\equiv1$, the unitary operator $U_{t_{k}}$ reduces to the one used in the Kaczmarz algorithm \cite{Shao}, which is similar to classical relaxed Kaczmarz algorithm.

\begin{table}[htbp]
\centering
\begin{tabular}{p{14cm}}
\caption{The quantum relaxed Kaczmarz algorithm (Algorithm $1$)}\\
\hline\hline
1. Select an arbitrary unit vector $x_{0}$, and prepare the state $|x_{0}\rangle$ with preparation time $O(\log{n})$. The initial state can be expressed as
\begin{equation*}
| X_{k} \rangle=\frac{||x_{k}||}{v_{k}}| 0 \rangle^{\otimes 3k}| 00 \rangle| x_{k} \rangle+| 0^\bot \rangle^{\otimes 3k+2}| \cdots \rangle
\end{equation*}
with $v_{0}=1$.\\
2. Prepare the state
\begin{equation*}
| Y_{k} \rangle=\beta_{t_k}| 0 \rangle| X_{k} \rangle+\gamma_{t_k}| 1 \rangle| 0 \rangle^{\otimes 3k}| 00 \rangle| a_{t_{k}} \rangle,
\end{equation*}
where $\beta^2_{t_k}=\frac{v^2_k}{v^2_k+b^2_{t_k}}$ and $\gamma^2_{t_k}=1-\beta^2_{t_k}$.\\
3. Applying the operation $(I_2^{\otimes(3k+1)}\otimes U_{t_{k}})SWAP_{1,3k+2}$ to $| Y_{k} \rangle$, we obtain
\begin{equation*}
|\widetilde{x}_{k+1}\rangle=\frac{||x_{k+1}||}{v_{k+1}}| 0 \rangle^{\otimes 3(k+1)}| x_{k+1} \rangle+| 0^\bot \rangle^{\otimes 3(k+1)}| \cdots \rangle.
\end{equation*}
4. Applying $SWAP_{1,3(k+1)+1}SWAP_{2,3(k+1)+2}$ to $| 00 \rangle|\widetilde{x}_{k+1}\rangle$, we get
\begin{equation*}
| X_{k+1} \rangle=\frac{||x_{k+1}||}{v_{k+1}}| 0 \rangle^{\otimes 3(k+1)}| 00 \rangle| x_{k+1} \rangle+| 0^\bot \rangle^{\otimes 3(k+1)+2}| \cdots \rangle.
\end{equation*}
5. Set $k=k+1$ and go to step 2 until convergence.\\
\hline\hline
\end{tabular}
\end{table}

To analyze the complexity of the algorithm, we assume that $| a_t \rangle$ can be prepared by the time of $O(\log{n})$. Let the complexity of obtaining $| X_{k} \rangle$ be $\tau_k$. Then $\tau_{k+1}=\tau_k+O(\log{n})$. Since $\tau_0=O(\log{n})$, the complexity for step $T$ is $O(T\log{n})$.

\subsection{Quantum relaxed column iteration algorithm}

We present now the quantum relaxed column iteration algorithm.
Without loss of generality, we can assume that $||c_t||=1$ for all $t$. Then the classical iteration formula of the relaxed column iteration (\ref{omega1}) and (\ref{omega2}) can be simplified to
\begin{eqnarray*}
x_{k+1}&=&x_{k}+\omega_{k}c_{t_{k}}^T r_{k}e_{t_{k}},\\
r_{k+1}&=&r_{k}-\omega_{k}c_{t_{k}}c_{t_{k}}^T r_{k}=(I_n-\omega_{k}c_{t_{k}}c_{t_{k}}^T)r_{k}.
\end{eqnarray*}
For any $c_t$ there exists an efficient unitary operator $S_t$ such that $\langle t |S_t=\langle c_t |$, $t=1,2,...,n$.
Accordingly, we obtain the quantum state corresponding to the iterative formula,
\begin{eqnarray*}
| x_{k+1} \rangle&\varpropto&||x_{k}||| x_{k} \rangle+\omega_{k}||r_k||| t_{k} \rangle\langle c_{t_k} | r_k \rangle\\
&=&||x_{k}||| x_{k} \rangle+\omega_{k}||r_k||| t_{k} \rangle\langle t_k |S_{t_k}| r_k \rangle,\\
| r_{k+1} \rangle&\varpropto&||r_k||(I_n-\omega_{k}| c_{t_{k}} \rangle \langle  c_{t_k} |)| r_k \rangle.
\end{eqnarray*}

We first construct the residual formula. Assume that the quantum state of the residual $r_{k}$ is included in
\begin{equation*}
| R_{k} \rangle=||r_k||| 0 \rangle^{\otimes 2k}| 00 \rangle| r_k \rangle+| 0^\bot \rangle^{\otimes 2(k+1)}| \cdots \rangle.
\end{equation*}
In order to get $| R_{k+1} \rangle$, we apply $(I_2^{\otimes 2k}\otimes U_{t_{k}})$ to $| R_{k} \rangle$. We get
\begin{eqnarray*}
(I_2^{\otimes 2k}\otimes U_{t_{k}})| R_{k} \rangle&=&| 0 \rangle^{\otimes 2(k+1)}||r_k||(I_n-\omega_{k}| c_{t_{k}} \rangle \langle  c_{t_k} |)| r_k \rangle+| 0^\bot \rangle^{\otimes 2(k+1)}| \cdots \rangle\\
&=&|\widetilde{r}_{k+1} \rangle,\\
| R_{k+1} \rangle&=&SWAP_{2,2k+4}SWAP_{1,2k+3}| 00 \rangle|\widetilde{r}_{k+1} \rangle\\
&=&||r_{k+1}||| 0 \rangle^{\otimes 2(k+1)}| 00 \rangle| r_{k+1} \rangle+| 0^\bot \rangle^{\otimes 2(k+2)}| \cdots \rangle,
\end{eqnarray*}
where
\[
U_{t_{k}}=\begin{bmatrix}
I_{n}-\omega_{k}| c_{t_{k}} \rangle \langle  c_{t_{k}} | & \sqrt{2\omega_{k}(1-\omega_{k})}| c_{t_{k}} \rangle \langle c_{t_{k}} | & \omega_{k}| c_{t_{k}} \rangle \langle  c_{t_{k}} | & 0\\
\sqrt{2\omega_{k}(1-\omega_{k})}| c_{t_{k}} \rangle \langle c_{t_{k}} | & 2\omega_{k}| c_{t_{k}} \rangle \langle c_{t_{k}} |-I_{n} & -\sqrt{2\omega_{k}(1-\omega_{k})}| c_{t_{k}} \rangle \langle c_{t_{k}} | & 0\\
\omega_{k}| c_{t_{k}} \rangle \langle  c_{t_{k}} | & -\sqrt{2\omega_{k}(1-\omega_{k})}| c_{t_{k}} \rangle \langle c_{t_{k}} | & I_{n}-\omega_{k}| c_{t_{k}} \rangle \langle  c_{t_{k}} | & 0\\
0&0&0&I_{n}
\end{bmatrix}.
\]

Next, consider the iterative process
\begin{equation*}
| x_{k+1}\rangle\varpropto||x_{k}||| x_{k} \rangle+\omega_{k}||r_k||| t_{k} \rangle\langle t_k |S_{t_k}| r_k \rangle.
\end{equation*}
Define $| \widetilde{X}_k\rangle=\frac{||x_{k}||}{v_k}| 0 \rangle^{\otimes 2}| x_{k} \rangle+| 0^\bot \rangle^{\otimes 2}| \cdots \rangle$ and $| \widetilde{R}_{k} \rangle=||r_k||| 0 \rangle^{\otimes 2} S_{t_k}| r_k \rangle+| 0^\bot \rangle^{\otimes 2}| \cdots \rangle$. We introduce ancilla qubits to prepare
$| \varphi_1 \rangle=\beta | 00 \rangle| \widetilde{X}_k\rangle +\gamma | 10 \rangle| \widetilde{R}_{k} \rangle$ with $\beta^2+\gamma^2=1$.
Denote
\[
W_{t_{k}}=\begin{bmatrix}
 \begin{array}{c|ccc}
 I_{n} &0 &0 &0\\ \hline
 0& & & \\
 0& & \huge{w_{t_k}} & \\
 0& & &
 \end{array}
\end{bmatrix},
\]
where
\[
w_{t_{k}}=\begin{bmatrix}
 I_{n}-\omega_{k}| t_{k} \rangle \langle t_{k} | & \omega_{k}| t_{k} \rangle \langle t_{k} | & \sqrt{2\omega_{k}(1-\omega_{k})}| t_{k} \rangle \langle t_{k} |\\
 \omega_{k}| t_{k} \rangle \langle  t_{k} | & I_{n}-\omega_{k}| t_{k} \rangle \langle t_{k} | & -\sqrt{2\omega_{k}(1-\omega_{k})}| t_{k} \rangle \langle t_{k} |\\
 \sqrt{2\omega_{k}(1-\omega_{k})}| t_{k} \rangle \langle t_{k} |&-\sqrt{2\omega_{k}(1-\omega_{k})}| t_{k} \rangle \langle t_{k} |& 2\omega_{k}| t_{k} \rangle \langle t_{k} |-I_{n}
\end{bmatrix}.
\]
Applying $W_{t_{k}}SWAP_{2,4}SWAP_{1,3}$ to $| \varphi_1 \rangle$, we obtain
\begin{eqnarray*}
|\varphi_2 \rangle&=&W_{t_{k}}SWAP_{2,4}SWAP_{1,3}| \varphi_1 \rangle\\
&=&W_{t_{k}}(| 00 \rangle\otimes(\beta\frac{||x_{k}||}{v_k}| 00 \rangle| x_{k} \rangle+\omega_{k}||r_k||| 01 \rangle S_{t_k}| r_k \rangle)+| 0^\bot \rangle^{\otimes 4}| \cdots \rangle)\\
&=&| 000 \rangle\otimes(\beta\frac{||x_{k}||}{v_k}| 0 \rangle| x_{k} \rangle+\gamma ||r_k||| 1 \rangle\omega_{k}| t_{k} \rangle \langle t_{k} |S_{t_k}| r_k \rangle)+| 0^\bot \rangle^{\otimes 4}| \cdots \rangle.
\end{eqnarray*}

In order to obtain the state $| x_{k+1} \rangle$, we apply the Givens Rotation
\[
G_{k}=\begin{bmatrix}
c&s\\
-s&c
\end{bmatrix},
\]
where $c^2+s^2=1$.
We obtain
\begin{eqnarray*}
| \varphi_3 \rangle&=&(I_2^{\otimes3}\otimes G_k \otimes I_n)| \varphi_2 \rangle\\
&=&| 0 \rangle^{\otimes 4}\otimes(\frac{c\beta||x_k||}{v_k}| x_{k} \rangle+s\gamma||r_k||\omega_{k}| t_{k} \rangle\langle t_k |S_{t_k}| r_k \rangle)+| 0^\bot \rangle^{\otimes 4}| \cdots \rangle.
\end{eqnarray*}
Choosing appropriate parameters $c$, $s$, $\beta$ and $\gamma$, we get
\begin{equation*}
\frac{c\beta||x_k||}{v_k}| x_{k} \rangle+s\gamma||r_k||\omega_{k}| t_{k} \rangle\langle t_k |S_{t_k}| r_k \rangle=||x_{k+1}||| x_{k+1} \rangle.
\end{equation*}
The detailed quantum relaxed column iteration algorithm is presented in Table II.
\begin{table}[htbp]
\centering
\begin{tabular}{p{14cm}}
\caption{The quantum relaxed column iteration algorithm (Algorithm 2)}
\\
\hline\hline
1. Select an arbitrary unit vector $x_{0}$ and prepare the state $|x_{0}\rangle$ within preparation time $O(\log{n})$. Assume that $r_0$ has unit norm and its quantum state is prepared in time $O(\log {n})$.
The initial states can be expressed as
\begin{eqnarray*}
| X_{k} \rangle&=&\frac{||x_{k}||}{k+1}| 0 \rangle^{\otimes 2(k+1)}| x_{k} \rangle+|  0^\bot \rangle^{\otimes 2(k+1)}| \cdots \rangle, \\
| R_{k} \rangle&=&||r_k||| 0 \rangle^{\otimes 2k}| 00 \rangle| r_k \rangle+| 0^\bot \rangle^{\otimes 2(k+1)}| \cdots \rangle,
\end{eqnarray*}
where $v_{k}=k+1$ and $r_0=b-Ax_0$.\\
2. Prepare the state
\begin{equation*}
| \psi \rangle=\sqrt{\frac{k+1}{k+2}}| 00 \rangle| X_{k} \rangle+\sqrt{\frac{1}{k+2}}| 10 \rangle(I^{\otimes {2(k+1)}}\otimes S_{t_k})| R_{k} \rangle.
\end{equation*}
3. Apply $(I_2^{\otimes2(k+1)+1}\otimes G_{k}\otimes I_n)(I_2^{\otimes2(k+1)}\otimes W_{t_k})SWAP_{2,2(k+1)+2}SWAP_{1,2(k+1)+1}$ to $| \psi \rangle$, we obtain $| X_{k+1} \rangle$.\\
4. Apply $SWAP_{2,2k+4}SWAP_{1,2k+3}$ to $| 00 \rangle|\widetilde{r}_{k+1} \rangle$ to obtain $| R_{k+1} \rangle $, where $|\widetilde{r}_{k+1} \rangle=(I_2^{\otimes 2k}\otimes U_{t_{k}})| R_{k} \rangle$.\\
5. Set $k=k+1$ and go to step 2 until convergence.\\
\hline\hline
\end{tabular}
\end{table}

In this algorithm, we select $c=\beta=\sqrt{\frac{k+1}{k+2}}$, $s=\gamma=\sqrt{\frac{1}{k+2}}$ and $v_k=k+1$ \cite{Shao}.
In step $2$, we prepare the state
\begin{eqnarray*}
| \psi \rangle&=&\sqrt{\frac{k+1}{k+2}}| 00 \rangle| X_{k} \rangle+\sqrt{\frac{1}{k+2}}| 10 \rangle(I^{\otimes {2(k+1)}}\otimes S_{t_k})| R_{k} \rangle\\
&=&\sqrt{\frac{k+1}{k+2}}| 00 \rangle(\frac{||x_{k}||}{k+1}| 0 \rangle^{\otimes 2k}| 00 \rangle| x_{k} \rangle+|  0^\bot \rangle^{\otimes 2(k+1)}| \cdots \rangle)\\
&\qquad&+\sqrt{\frac{1}{k+2}}| 10 \rangle(||r_k||| 0 \rangle^{\otimes 2k}| 00 \rangle S_{t_k}| r_k \rangle+| 0^\bot \rangle^{\otimes 2(k+1)}| \cdots \rangle).
\end{eqnarray*}

In step $3$, we apply $(I_2^{\otimes2(k+1)+1}\otimes G_{k}\otimes I_n)(I_2^{\otimes2(k+1)}\otimes W_{t_k})SWAP_{2,2(k+1)+2}SWAP_{1,2(k+1)+1}$ to $| \psi \rangle$. For the application of $SWAP_{2,2(k+1)+2}SWAP_{1,2(k+1)+1}$ to $| \psi \rangle$, we have
\begin{eqnarray*}
| \psi_1 \rangle&=&\sqrt{\frac{k+1}{k+2}}(\frac{||x_{k}||}{k+1}| 0 \rangle^{\otimes 2(k+1)}| 00 \rangle| x_{k} \rangle+|  0^\bot \rangle^{\otimes 2(k+1)}| 00 \rangle| \cdots \rangle)\\
&\qquad&+\sqrt{\frac{1}{k+2}}(||r_k||| 0 \rangle^{\otimes 2(k+1)}| 10 \rangle S_{t_k}| r_k \rangle+| 0^\bot \rangle^{\otimes 2(k+1)}| 10 \rangle| \cdots \rangle)\\
&=&| 0 \rangle^{\otimes 2(k+1)}\otimes(\sqrt{\frac{k+1}{k+2}}\frac{||x_{k}||}{k+1}| 00 \rangle| x_{k} \rangle+\sqrt{\frac{1}{k+2}}||r_k||| 10 \rangle S_{t_k}| r_k \rangle)+| 0^\bot \rangle^{\otimes 2(k+2)}| \cdots \rangle.
\end{eqnarray*}
Applying $I_2^{\otimes2(k+1)}\otimes W_{t_k}$ to $| \psi_1 \rangle$ we get
\begin{eqnarray*}
|\psi_2\rangle&=&I_2^{\otimes2(k+1)}\otimes W_{t_k}| \psi_1 \rangle\\
&=&|0\rangle^{\otimes 2(k+1)}\otimes(\sqrt{\frac{k+1}{k+2}}\frac{||x_{k}||}{k+1}| 00 \rangle|| x_{k} \rangle\\
&\qquad&+\sqrt{\frac{1}{k+2}}||r_k||(| 01 \rangle \omega_{k}| t_{k} \rangle \langle t_{k} |S_{t_k}| r_k \rangle+| 0^\bot \rangle^{\otimes 2}| \cdots \rangle))\\
&\qquad&+| 0^\bot \rangle^{\otimes 2(k+2)}| \cdots \rangle\\
&=&|0\rangle^{\otimes 2(k+1)+1}(\sqrt{\frac{k+1}{k+2}}\frac{||x_{k}||}{k+1}| 0 \rangle| x_{k} \rangle+\sqrt{\frac{1}{k+2}}||r_k||| 1 \rangle \omega_{k}| t_{k} \rangle \langle t_{k} |S_{t_k}| r_k \rangle)\\
&\qquad&+|0^\bot\rangle^{\otimes 2(k+2)}| \cdots \rangle.
\end{eqnarray*}
Finally, applying $I_2^{\otimes2(k+1)+1}\otimes G_{k}\otimes I_n$ to $| \psi_2 \rangle$, we obtain $| X_{k+1} \rangle$,
\begin{eqnarray*}
| X_{k+1} \rangle
&=&| 0 \rangle^{\otimes 2(k+2)}(\frac{k+1}{k+2}\frac{||x_{k}||}{k+1}| x_{k} \rangle+\frac{||r_k||}{k+2} \omega_{k}| t_{k} \rangle \langle t_{k} |S_{t_k}| r_k \rangle)+|0^\bot \rangle^{\otimes 2(k+2)}| \cdots \rangle\\
&=&\frac{||x_{k+1}||}{k+2}| 0 \rangle^{\otimes 2(k+2)}| x_{k+1} \rangle+| 0^\bot \rangle^{\otimes 2(k+2)}| \cdots \rangle.
\end{eqnarray*}

For iterative step $T$, Algorithm 2 is the same as the Algorithm 1. The complexity of the relaxation column iteration algorithm is $O(T\log{n})$. Therefore, exponential acceleration can be achieved for quantum relaxed column iteration method based on block-encoding.

Similarly, when we set $\omega_{k}\equiv1$ and adjust the algorithm appropriately, the quantum relaxed column iteration algorithm can reduce to the quantum column iteration algorithm. Moreover, for appropriate $x_0$, $r_0$ has always unit norm. Otherwise, by selecting a parameter $\delta$ with $\delta||r_0||\leq1$, one can transform $| R_{0} \rangle$ to $| \hat{R_{0}} \rangle=\delta||r_0||| 00 \rangle| r_0 \rangle+| 11 \rangle| \cdots \rangle$ to make sure algorithm 2 works \cite{Shao}.

\section{Numerical Analysis}

In this section, we demonstrate the feasibility of our algorithms by exhibiting the following examples. For simplicity, we just only iterate our algorithms twice in the examples.

\subsection{An example for Algorithm $1$}

We use the quantum relaxed Kaczmarz algorithm to solve the following system of linear equations,
\begin{displaymath}
\left\{ \begin{array}{ll}
\frac{1}{\sqrt{2}}x_{1}+\frac{1}{\sqrt{2}}x_{2}=2\sqrt{2},\\
\frac{1}{\sqrt{2}}x_{1}-\frac{1}{\sqrt{2}}x_{2}=\sqrt{2},\\
\end{array} \right.
\end{displaymath}
with $a_{1}={(1/\sqrt{2}, 1/\sqrt{2})}^{T}$ and $a_{2}={(1/\sqrt{2}, -1/\sqrt{2})}^{T}$.

In step $1$ and step $2$, we select the unit vector $x_{0}={(1, 0)}^{T}$ and $v_{0}=1$. We have
\begin{eqnarray*}
| X_{0} \rangle&=&\frac{||x_{0}||}{v_{0}}| 00 \rangle| x_{0} \rangle,\\
| Y_{0} \rangle&=&\beta_{t_0}| 0 \rangle| X_{0} \rangle+\gamma_{t_0}| 1 \rangle| 00 \rangle| a_{t_{0}} \rangle\\
&=&\beta_{1_0}| 0 \rangle\frac{||x_{0}||}{v_{0}}| 00 \rangle| x_{0} \rangle+\gamma_{1_0}| 1 \rangle| 00 \rangle| a_{1_{0}} \rangle,
\end{eqnarray*}
where $t=1$ and $\lambda_{0}=\frac{1}{3}$.

In step $3$, we apply $(I_2\otimes U_{1_{0}})SWAP_{1,2}$ to $| Y_{0} \rangle$ and obtain
\begin{eqnarray*}
(I_2\otimes U_{1_{0}})SWAP_{1,2}|Y_{0}\rangle&=&(I_2\otimes U_{1_{0}})(\beta_{1_0}\frac{||x_{0}||}{v_{0}}| 0 \rangle| 00 \rangle| x_{0} \rangle+\gamma_{1_0}| 0 \rangle| 10 \rangle| a_{1_{0}} \rangle)\\
&=&|000\rangle\otimes[\beta_{1_0}\frac{||x_{0}||}{v_{0}}(I_{2}-\lambda_{0}|a_{1_{0}}\rangle \langle a_{1_{0}}|)| x_{0} \rangle+\gamma_{1_0}\lambda_{0}| a_{1_{0}} \rangle]\\
&\qquad&+| 0 \rangle\otimes(| 01 \rangle| \cdots \rangle+| 10 \rangle| \cdots \rangle).
\end{eqnarray*}

Due to $\beta^2_{t_k}=\frac{v^2_k}{v^2_k+b^2_{t_k}}$, $\gamma_{t_k}=\frac{\beta_{t_k}b_{t_k}}{v_{k}}$ and $\frac{\beta_{t_k}}{v_{k}}=\frac{1}{v_{k+1}}$, we compute
\begin{eqnarray*}
|\widetilde{x}_{1}\rangle&=&| 0 \rangle^{\otimes 3}[\frac{\beta_{1_0}}{v_{0}}(||x_{0}||(I_{2}-\lambda_{0}|a_{1_{0}}\rangle \langle a_{1_{0}}|)| x_{0} \rangle)+b_{1_0}\lambda_{0}| a_{1_{0}}\rangle]\\
&\qquad&+| 0 \rangle\otimes(| 01 \rangle| \cdots \rangle+| 10 \rangle| \cdots \rangle)\\
&=&\frac{||x_{1}||}{v_{1}}| 0 \rangle^{\otimes 3}| x_{1} \rangle+| 0^\bot \rangle^{\otimes 3}| \cdots \rangle,
\end{eqnarray*}
where $\beta_{1_0}=1/3$, $\frac{1}{v_{1}}=\frac{\beta_{1_0}}{v_{0}}=1/3$, $||x_1||=\frac{\sqrt{10}}{2}$ and $| x_{1} \rangle=(\frac{3}{10}\sqrt{10}, \frac{\sqrt{10}}{10})^{T}$.
Applying $SWAP_{1,4}SWAP_{2,5}$ to $| 00 \rangle|\widetilde{x}_{1}\rangle$ we obtain
\begin{equation*}
| X_{1} \rangle=\frac{||x_{1}||}{v_{1}}| 0 \rangle^{\otimes 3}| 00 \rangle| x_{1} \rangle+| 0^\bot \rangle^{\otimes 5}| \cdots \rangle.
\end{equation*}

Next, we perform the second iteration by setting
\begin{equation*}
| Y_{1} \rangle=\beta_{t_1}| 0 \rangle| X_{1} \rangle+\gamma_{t_1}| 1 \rangle| 0 \rangle^{\otimes 3}| 00 \rangle| a_{t_{1}} \rangle.
\end{equation*}
Choosing $t=2$ and $\lambda_{1}=1$, we have
\begin{eqnarray*}
| Y_{1} \rangle=\beta_{2_1}| 0 \rangle(\frac{||x_{1}||}{v_{1}}| 0 \rangle^{\otimes 3}| 00 \rangle| x_{1} \rangle+| 0^\bot \rangle^{\otimes 5}| \cdots \rangle)+\gamma_{2_1}| 1 \rangle| 0 \rangle^{\otimes 3}| 00 \rangle| a_{2_{1}} \rangle.
\end{eqnarray*}

In step $3$, applying $(I_2^{\otimes4}\otimes U_{1_{0}})SWAP_{1,5}$ to $| Y_{1} \rangle$ we get
\begin{eqnarray*}
(I_2^{\otimes4}&\otimes& U_{1_{0}})SWAP_{1,5}| Y_{1} \rangle\\
&=&(I_2^{\otimes4}\otimes U_{1_{0}})(\beta_{2_1}\frac{||x_{1}||}{v_{1}}| 0 \rangle^{\otimes 4}| 00 \rangle| x_{1} \rangle+\gamma_{2_1}| 0 \rangle^{\otimes 4}| 10 \rangle| a_{2_{1}} \rangle+| 0^\bot \rangle^{\otimes 4}| 00 \rangle| \cdots \rangle)\\
&=&|0\rangle^{\otimes 4}| 00\rangle[\beta_{2_1}\frac{||x_{1}||}{v_{1}}(I_{2}-\lambda_{1}|a_{2_{1}}\rangle \langle a_{2_{1}}|)| x_{1} \rangle+\gamma_{2_1}\lambda_{1}| a_{2_{1}} \rangle]+| 0^\bot \rangle^{\otimes 6}| \cdots \rangle.
\end{eqnarray*}
From $\beta^2_{t_k}=\frac{v^2_k}{v^2_k+b^2_{t_k}}$, $\gamma_{t_k}=\frac{\beta_{t_k}b_{t_k}}{v_{k}}$ and $\frac{\beta_{t_k}}{v_{k}}=\frac{1}{v_{k+1}}$, we calculate that
\begin{align*}
|\widetilde{x}_{2}\rangle&=| 0 \rangle^{\otimes 6}[\frac{\beta_{2_1}}{v_{1}}(||x_{1}||(I_{2}-\lambda_{1}|a_{2_{1}}\rangle \langle a_{2_{1}}|)| x_{1} \rangle)+b_{2_1}\lambda_{1}| a_{2_{1}}\rangle]+| 10 \rangle| \cdots \rangle)+| 0^\bot \rangle^{\otimes 6}| \cdots \rangle\\
&=\frac{||x_{2}||}{v_{2}}| 0 \rangle^{\otimes 6}| x_{1} \rangle+| 0^\bot \rangle^{\otimes 6}| \cdots \rangle,
\end{align*}
where $\beta_{2_1}=\frac{3\sqrt{11}}{11}$, $\frac{1}{v_{2}}=\frac{\beta_{2_1}}{v_{1}}=\frac{\sqrt{11}}{11}$, $||x_2||=2$ and $| x_{2} \rangle=(1,0)^{T}$.
Applying $SWAP_{1,7}SWAP_{2,8}$ to $| 00 \rangle|\widetilde{x}_{2}\rangle$, we obtain
$$| X_{2} \rangle=\frac{||x_{2}||}{v_{2}}| 0 \rangle^{\otimes 6}| 00 \rangle| x_{2} \rangle+| 0^\bot \rangle^{\otimes 8}| \cdots \rangle.$$

Based on $|X_2\rangle$, one can repeat algorithm $1$ until the result converges to the desired precision.

\subsection{An example of algorithm $2$}

We apply the Algorithm $2$ to solve the following set of linear equations,
\begin{displaymath}
\left\{ \begin{array}{ll}
-\frac{1}{\sqrt{2}}x_{1}+\frac{1}{\sqrt{2}}x_{2}=\sqrt{2},\\
-\frac{1}{\sqrt{2}}x_{1}-\frac{1}{\sqrt{2}}x_{2}=0,\\
\end{array} \right.
\end{displaymath}
where we have set $c_{1}={(-\frac{1}{\sqrt{2}}, -\frac{1}{\sqrt{2}})}^{T}$ and $c_{2}={(1/\sqrt{2}, -1/\sqrt{2})}^{T}$.

Selecting the unit vector $x_{0}={(0, 1)}^{T}$ and setting $r_0=b-Ax_0={(1/\sqrt2, 1/\sqrt2)}^{T}$, we produce
\begin{eqnarray*}
| R_{0} \rangle&=&||r_0||| 00 \rangle| r_0 \rangle,\\
| X_{0} \rangle&=&\frac{||x_{0}||}{0+1}| 0 \rangle^{\otimes 2(0+1)}| x_{0} \rangle,\\
| \psi \rangle&=&\sqrt{\frac{0+1}{0+2}}| 00 \rangle| X_{0} \rangle+\sqrt{\frac{1}{0+2}}| 10 \rangle(I^{\otimes {2}}\otimes S_{t_0})| R_{0} \rangle.
\end{eqnarray*}
Applying $(I_2^{\otimes3}\otimes G_{0}\otimes I_2)(I_2^{\otimes2}\otimes W_{t_0})SWAP_{2,4}SWAP_{1,3}$ to $| \psi \rangle$, we compute
\begin{eqnarray*}
(I_2^{\otimes3}&\otimes&G_{0}\otimes I_2)(I_2^{\otimes2}\otimes W_{t_0})SWAP_{2,4}SWAP_{1,3}| \psi \rangle\\
&=&(I_2^{\otimes3}\otimes G_{0}\otimes I_2)(I_2^{\otimes2}\otimes W_{t_0})[\sqrt{\frac{0+1}{0+2}}\frac{||x_{0}||}{0+1}| 00 \rangle| 00 \rangle| x_{0} \rangle+\sqrt{\frac{1}{0+2}}||r_0||| 00 \rangle| 10 \rangle S_{t_0}| r_0 \rangle]\\
&=&(I_2^{\otimes3}\otimes G_{0}\otimes I_2)[| 00 \rangle(\sqrt{\frac{0+1}{0+2}}\frac{||x_{0}||}{0+1}| 00 \rangle| x_{0} \rangle\\
&\qquad&+\sqrt{\frac{1}{0+2}}||r_0||| 01 \rangle\omega_{k}| t_{0} \rangle \langle t_{0} | S_{t_0}| r_0 \rangle+| 10 \rangle | \cdots\rangle+| 11 \rangle | \cdots\rangle)]\\
&=&(I_2^{\otimes3}\otimes G_{0}\otimes I_2)[| 000 \rangle(\sqrt{\frac{0+1}{0+2}}\frac{||x_{0}||}{0+1}| 0 \rangle| x_{0} \rangle\\
&\qquad&+\sqrt{\frac{1}{0+2}}||r_0||| 1 \rangle\omega_{0}| t_{0} \rangle \langle t_{0} | S_{t_0}| r_0 \rangle)+| 0010 \rangle | \cdots\rangle+| 0011 \rangle | \cdots\rangle\\
&=&| 0000 \rangle(\sqrt{\frac{0+1}{0+2}}\frac{||x_{0}||}{0+1}c| x_{0} \rangle+\sqrt{\frac{1}{0+2}}||r_0||s\omega_{0}| t_{0} \rangle \langle t_{0} | S_{t_0}| r_0 \rangle)+| 0^\bot \rangle^{\otimes 4}| \cdots \rangle.
\end{eqnarray*}

Taking $c=\sqrt{\frac{0+1}{0+2}}$, $s=\sqrt{\frac{1}{0+2}}$, $\omega_{0}=\frac{1}{2}$ and $t=1$, we get
\begin{equation*}
| X_{1} \rangle=\frac{||x_{1}||}{2}| 0 \rangle^{\otimes 4}| x_{1} \rangle+| 0^\bot \rangle^{\otimes 4}| \cdots \rangle,
\end{equation*}
where $||x_1||=\frac{\sqrt5}{2}$ and $|x_1\rangle={(-\frac{1}{\sqrt5}, \frac{2}{\sqrt5})}^{T}$.
We have
\begin{eqnarray*}
|\widetilde{r}_{1} \rangle&=& U_{1_{0}}| R_{0} \rangle\\
&=&||r_0||| 00 \rangle (I_{2}-\omega_{0}| c_{1_{0}} \rangle \langle  c_{1_{0}} |)| r_0 \rangle+| 0^\bot \rangle^{\otimes 2}| \cdots \rangle\\
&=&| 00 \rangle||r_1||| r_1 \rangle+| 0^\bot \rangle^{\otimes 2}| \cdots \rangle,
\end{eqnarray*}
where $||r_1||=\frac{1}{2}$ and $|r_1\rangle={(\frac{1}{\sqrt2}, \frac{1}{\sqrt2})}^{T}$.
Then applying $SWAP_{2,4}SWAP_{1,3}$ to $| 00 \rangle|\widetilde{r}_{1} \rangle$, one obtains
\begin{equation*}
|R_{1} \rangle=|00\rangle||r_1||| 00 \rangle| r_1 \rangle+| 0^\bot \rangle^{\otimes 4}| \cdots \rangle.
\end{equation*}

We then set
\begin{eqnarray*}
|\psi\rangle&=&\sqrt{\frac{1+1}{1+2}}| 00 \rangle| X_{1} \rangle+\sqrt{\frac{1+1}{1+2}}| 10 \rangle(I^{\otimes {4}}\otimes S_{t_1})| R_{1} \rangle\\
&=&\sqrt{\frac{1+1}{1+2}}| 00 \rangle(\frac{||x_{1}||}{2}| 0 \rangle^{\otimes 4}| x_{1} \rangle+| 0^\bot \rangle^{\otimes 4}| \cdots \rangle)\\
&\qquad&+\sqrt{\frac{1+1}{1+2}}| 10 \rangle(| 0 \rangle^{\otimes 4}||r_1|||S_{t_1}| r_1 \rangle+| 0^\bot \rangle^{\otimes 4}| \cdots \rangle).
\end{eqnarray*}
Applying $(I_2^{\otimes6}\otimes G_{1}\otimes I_2)(I_2^{\otimes4}\otimes W_{t_1})SWAP_{2,6}SWAP_{1,5}$ to $| \psi \rangle$, we obtain $| X_{2} \rangle$,
\begin{eqnarray*}
(I_2^{\otimes6}&\otimes& G_{1}\otimes I_2)(I_2^{\otimes4}\otimes W_{t_1})SWAP_{2,6}SWAP_{1,5}| \psi \rangle\\
&=&(I_2^{\otimes6}\otimes G_{1}\otimes I_2)(I_2^{\otimes4}\otimes W_{t_1})[| 0 \rangle^{\otimes 4}(\sqrt{\frac{1+1}{1+2}}| 00 \rangle\frac{||x_{1}||}{2}| x_{1} \rangle\\
&\qquad&+\sqrt{\frac{1+1}{1+2}}| 10 \rangle||r_1||S_{t_1}| r_1 \rangle)+| 0^\bot \rangle^{\otimes 4}| \cdots \rangle]\\
&=&(I_2^{\otimes6}\otimes G_{1}\otimes I_2)[| 0 \rangle^{\otimes 5}(\sqrt{\frac{1+1}{1+2}}| 0 \rangle\frac{||x_{1}||}{2}| x_{1} \rangle\\
&\qquad&+\sqrt{\frac{1+1}{1+2}}||r_1||| 1 \rangle \omega_{1}| t_{1} \rangle \langle t_{1} | S_{t_1}| r_1 \rangle))+| 0^\bot \rangle^{\otimes 5}| \cdots \rangle]\\
&=&| 0 \rangle^{\otimes 6}(\sqrt{\frac{1+1}{1+2}}\frac{||x_{1}||}{2}c| x_{1} \rangle+\sqrt{\frac{1+1}{1+2}}||r_1||s\omega_{1}| t_{1} \rangle \langle t_{1} | S_{t_1}| r_1 \rangle))+| 0^\bot \rangle^{\otimes 6}| \cdots \rangle.
\end{eqnarray*}

Set $c=\sqrt{\frac{1+1}{1+2}}$, $s=\sqrt{\frac{1}{1+2}}$, $\omega_{1}=1$ and $t=1$. We obtain
\begin{equation*}
| X_{2} \rangle=\frac{||x_{2}||}{3}| 0 \rangle^{\otimes 6}| x_{2} \rangle+| 0^\bot \rangle^{\otimes 6}| \cdots \rangle,
\end{equation*}
where $||x_2||=\sqrt2$ and $|x_2\rangle={(-\frac{1}{\sqrt2}, \frac{1}{\sqrt2})}^{T}$.
And we compute
\begin{eqnarray*}
|\widetilde{r}_{2} \rangle&=&(I_2^{\otimes2}\otimes U_{1_{1}})| R_{1} \rangle\\
&=&||r_1||| 0 \rangle^{\otimes 4} (I_{2}-\omega_{1}| c_{1_{1}} \rangle \langle  c_{1_{1}} |)| r_1 \rangle+| 0^\bot \rangle^{\otimes 4}| \cdots \rangle.
\end{eqnarray*}
Applying $SWAP_{2,6}SWAP_{1,5}$ to $| 00 \rangle|\widetilde{r}_{2} \rangle$, one obtains
\begin{equation*}
| R_{2} \rangle=| 0 \rangle^{\otimes 4}||r_2||| 00 \rangle| r_2 \rangle+| 0^\bot \rangle^{\otimes 6}| \cdots \rangle,\\
\end{equation*}
where $||r_2||=0$ and $|r_{2}\rangle={(0, 0)}^{T}$.

Through the above iterations, we have already got the exact solution of the linear system of equations.

\section{Conclusion}

We have presented quantum relaxed row and column iteration methods based on block-encoding and implemented proof-of-principle numerical verification. With the assumption that quantum states can be prepared efficiently, we have constructed the unitary operations for the relaxed row and column algorithms. Independence on the detailed Hamiltonian simulations and quantum phase estimations, the complexity of the algorithms is linear with the iterative steps, and achieves an exponential acceleration in the dimension $n$. Furthermore, choosing an expedient relaxation factor will significantly accelerate the convergence of iteration. In both algorithms, we can specify the maximum number of iterations. However, the maximum number of iterations is an empirical choice with uncertainty. We can consider a more appropriate quantum method for the optimal stopping time, which is an important future direction.

\noindent{\bf Acknowledgments and Data Availability Statements}\, \, This work is supported by the Fundamental Research Funds for the Central Universities 22CX03005A, the Shandong Provincial Natural Science Foundation for Quantum Science No. ZR2020LLZ003, ZR2021LLZ002, Shenzhen Institute for Quantum Science and Engineering, Southern University of Science and Technology (Grant No. SIQSE202001), Beijing Natural Science Foundation (Z190005), the Academician Innovation Platform of Hainan Province, Academy for Multidisciplinary Studies, Capital Normal University, and NSFC No. 12075159, 12171044. All data generated or analysed during this study are included in this published article.

\maketitle 
{\small \begin{thebibliography}{99}

\bibitem{Shor}P. W. Shor, Algorithms for quantum computation: discrete logarithms and factoring, Proc. of the 35th FOCS (IEEE, New York,1994), pp. 124-134.
\bibitem{Grover}K. Grover, Quantum Mechanics Helps in Searching for a Needle in a Haystack, Phys. Rev. Lett. 79(2): 325.1997.  arXiv:quant-ph/9706033.
\bibitem{HHL}A. W. Harrow, A. Hassidim, and S. Lloyd, Quantum Algorithm for Linear Systems of Equations, Phys. Rev. Lett. 103, 150502 (2009).
\bibitem{QML}J. Biamonte, P. Wittek, , N. Pancotti, P. Rebentros, N. Wiebe, and S. Lloyd, Quantum machine learning, Nature 549, 195C202 (2017).
\bibitem{QSVM}P. Rebentrost, M. Mohseni, and S. Lloyd, Quantum Support Vector Machine for Big Data Classification, Phys. Rev. Lett. 113, 130503  (2014).
\bibitem{FNN}K. H. Wan, O. Dahlsten, H. Kristjnsson, R. Gardner and M.S. Kim, Quantum generalisation of feedforward neural networks, arXiv:1612.01045 (2016).
\bibitem{NISQ}J. Preskill, Quantum computing in the NISQ era and beyond, Quantum 2, 79, arXiv:1801.00862 (2018).
\bibitem{VQA}A. Peruzzo, J. McClean, P. Shadbolt, M.-H. Yung, X.-Q. Zhou, P. J. Love, A. Aspuru-Guzik, and J. L. O'Brien, A variational eigenvalue solver on a photonic quantum processor, Nat Commun 5, 4213 (2014).
\bibitem{Kaczmarz}S. Kaczmarz, Classe des Sciences Math$\acute{e}$matiques et Naturelles. S$\acute{e}$rie A, Sciences Math$\acute{e}$matiques 35, 355 (1937).
\bibitem{Gordon}R.Gordon, R. Bender, and G. Herman, Algebraic reconstruction techniques (ART) for the three-dimensional electron miscroscopy and X-ray photography, J. Theor. Biol., 29, pp. 471-481 (1970).
\bibitem{CD}D. Leventhal and A. S. Lewis, Randomized Methods for Linear Constraints: Convergence Rates and Conditioning, Math. Oper. Res. 35, 641 (2010).
\bibitem{iteration1}P. Rebentrost, M. Schuld, L. Wossnig, F. Petruccione, and S.Lloyd, Quantum gradient descent and Newton's method for constrained
polynomial optimization, New J. Phys. 21, 073023 (2019).
\bibitem{iteration2} I. Kerenidis and A. Prakash, A Quantum Interior Point Method for LPs and SDPs, arXiv:1808.09266v1 (2018).
\bibitem{CKS}A. M. Childs, R. Kothari, and R. D. Somma, Quantum Algorithm for Systems of Linear Equations with Exponentially Improved Dependence on Precision, SIAM J. Comput. 46, 1920 (2017).
\bibitem{WZP}L. Wossnig, Z. Zhao, and A. Prakash, Quantum Linear System Algorithm for Dense Matrices, Phys. Rev. Lett. 120, 050502 (2018).
\bibitem{CGJ}S. Chakraborty, A. Gily$\acute{e}$n, and S. Jeffery, in 46th Interna tional Colloquium on Automata, Languages, and Program ming (ICALP 2019), Leibniz International Proceedings in In formatics (LIPIcs), edited by C. Baier, I. Chatzigiannakis, P. Flocchini, and S. Leonardi (Schloss Dagstuhl-Leibniz Zentrum fuer Informatik, Dagstuhl, Germany, 2019), Vol. 132, pp. 33:1-33:14.
\bibitem{Shao}C.P. Shao, H. Xiang, Row and column iteration methods to solve linear systems on a quantum computer, Physical Review A. 101, 022322 (2020).
\bibitem{Zuo}Q. Zuo, C.P. Shao, N.C. Wu, H. Xiang, An Extended Row and Column Method for Solving Linear Systems on a Quantum Computer. Int J Theor Phys (2021).
\bibitem{relaxed kaczmarz}P.P.B. Eggermont, G.T. Herman, A.Lent, Iterative algorithms for larg partitioned linear systems, with appications to image reconstruction, Linear Algebra Appl. 40(1981) 37-67.
\bibitem{relaxed cd1}T. Elfving, P.C. Hansen,and T. Nikazad,Convergence analysis for column-action methods in image reconstruction, Numer. Algor. 74, 905 (2017).
\bibitem{relaxed cd2}Kui Du, Xiaohui Sun, A doubly stochastic block Gauss-Seidel algorithm for solving linear equations, arXiv:1912.13291 (2020).
\bibitem{Whitney}T. M. Whitney and R. K. Meany, Two algorithms related to the method of steepest descent, SIAM J. Numer. Anal., 4 (1967) 109-118.
\bibitem{qram}V. Giovannetti, S. Lloyd, and L. Maccone, Quantum Random Access Memory, Phys. Rev. Lett. 100, 160501 (2008).
\bibitem{cases}L. Grover and T. Rudolph, Creating superpositions that correspond to efficiently integrable probability distributions,  arXiv:quant-ph/0208112 (2002).
\end{thebibliography}}
\end{document}